\documentclass[aps,prb,reprint,showpacs,superscriptaddress]{revtex4-1}
\usepackage{amssymb,amsmath,amsbsy,graphicx,hyperref,xcolor,csquotes,braket}

\begin{document}
\title{Magnetoelectronic properties of normal and skewed phosphorene nanoribbons}

\author{Vladimir V. Arsoski}\email{vladimir.arsoski@etf.bg.ac.rs}
\affiliation{School of Electrical Engineering, University of Belgrade, P.O. Box
3554, 11120 Belgrade, Serbia}
\author{Marko M. Gruji\'c}\email{marko.grujic@etf.bg.ac.rs}
\affiliation{School of Electrical Engineering, University of Belgrade, P.O. Box
3554, 11120 Belgrade, Serbia} \affiliation{Department of Physics, University of
Antwerp, Groenenborgerlaan 171, B-2020 Antwerp, Belgium}
\author{Nemanja A. \v{C}ukari\'c}\email{nemanja.cukaric@etf.bg.ac.rs}
\affiliation{School of Electrical Engineering, University of Belgrade, P.O. Box
3554, 11120 Belgrade, Serbia} \affiliation{Department of Physics, University of
Antwerp, Groenenborgerlaan 171, B-2020 Antwerp, Belgium}
\author{Milan \v{Z}. Tadi\'c}\email{milan.tadic@etf.bg.ac.rs}
\affiliation{School of Electrical Engineering, University of Belgrade, P.O. Box
3554, 11120 Belgrade, Serbia}
\author{Fran\c{c}ois M. Peeters}\email{francois.peeters@uantwerpen.be}
\affiliation{Department of Physics, University of Antwerp, Groenenborgerlaan
171, B-2020 Antwerp, Belgium}

\begin{abstract}
The energy spectrum and eigenstates of single-layer black phosphorous nanoribbons in the presence of perpendicular magnetic field and in-plane transverse electric field are investigated by means of a tight-binding method and the effect of different types of edges is analytically examined. A description based on a new continuous model is proposed by expansion of the tight-binding model in the long-wavelength approximation. The wavefunctions corresponding to the flatband part of the spectrum are obtained analytically and are shown to  approach agree well with the numerical results from the tight-binding method. Analytical expressions for the critical magnetic field at which Landau levels are formed and the ranges of wavenumbers in the dispersionless flat-band segments in the energy spectra are derived. We examine the evolution of the Landau levels when an in-plane lateral electric field is applied and determine analytically how the edge states shift with magnetic field.
\end{abstract}

 \maketitle

\section{Introduction}\label{I}

Phosphorene, a single layer of black phosphorus, is a recently isolated
material,\cite{liu14,lu14} providing some extraordinary advantages over other
two-dimensional (2D) materials. Most importantly, it has a large band gap of
about $1.84$ eV,\cite{rudenko15} thus circumventing the main drawback of
graphene with its zero band gap. Additionally, because of the puckered structure \cite{liu14,fei14,aierken15} 
it has highly anisotropic electronic and thermal properties,
allowing for the emergence of new functionalities. On top of this, higher
carrier mobilities can be achieved as compared to those 2D materials
consisting of transition metal dichalcogenides.\cite{qiao14}

However, phosphorene also has two major downsides: i) the quality of the
samples degrades with time when exposed to air,\cite{favron14} and ii) unlike
few-layer phosphorene the band gap is not electrically tunable. The first issue 
can be evaded successfully by for instance encapsulating
phosphorene with hexagonal boron nitride, which can has the added benefit of
increasing the crystal quality and carrier mobility, as in the case of
graphene.\cite{cao15,li15,chen15} Besides strain,\cite{rodin14,cakir14} fashioning 
phosphorene into nanoribbons can modify the band gap through the quantum 
confinement effect. Indeed, several theoretical papers have studied 
various types of phosphorene nanoribbons, and all of them show 
potentials for certain applications.\cite{ezawa14,sisakht15,grujic16} Note that
due to the anisotropy of phosphorene there are two distinct types of zigzag (ZZ) and
armchair (AC) edges, normal and skewed, with skewed edges intersecting the ridges of
phosphorus atoms at a sharp angle.\cite{grujic16} Intriguingly, normal zigzag
and skewed armchair (sAC) nanoribbons are metallic, while skewed zigzag (sZZ) and normal
armchair nanoribbons are insulating, and all of them have an electrically
tunable band structure, featuring metal-insulator transitions for a
sufficiently strong electric field.

While it was predicted that "bulk phosphorene" displays linearly dispersing
Landau levels (LLs) for low-energy quasiparticles in low magnetic
fields,\cite{pereira15,ostahie15,zhou15} which was confirmed by recent
experiments,\cite{li15,chen15} the results concerning the magnetic response of
phosphorene nanoribbons are relatively scarce. In Ref. ~[\onlinecite{zhou15}] the
formation of LLs in normal nanoribbons was reported, while in
Ref. ~[\onlinecite{ostahie15}] the impact of magnetic field on the quasi-flat
bands (QFBs) of zigzag nanoribbons were investigated. The purpose of this paper
is to examine the influence of magnetic field on phosphorene nanoribbons in
more details. In particular, we will examine the band structure of phosphorene
nanoribbons with various edge types (including skewed). We will discuss the impact of
dispersion anisotropy on LL formation from both the qualitative and quantitative
point of view. We will also briefly consider the effect of in-plane electric field on the LLs.

\section{Theoretical model}

\begin{figure} \centering
\includegraphics[width=8.6cm]{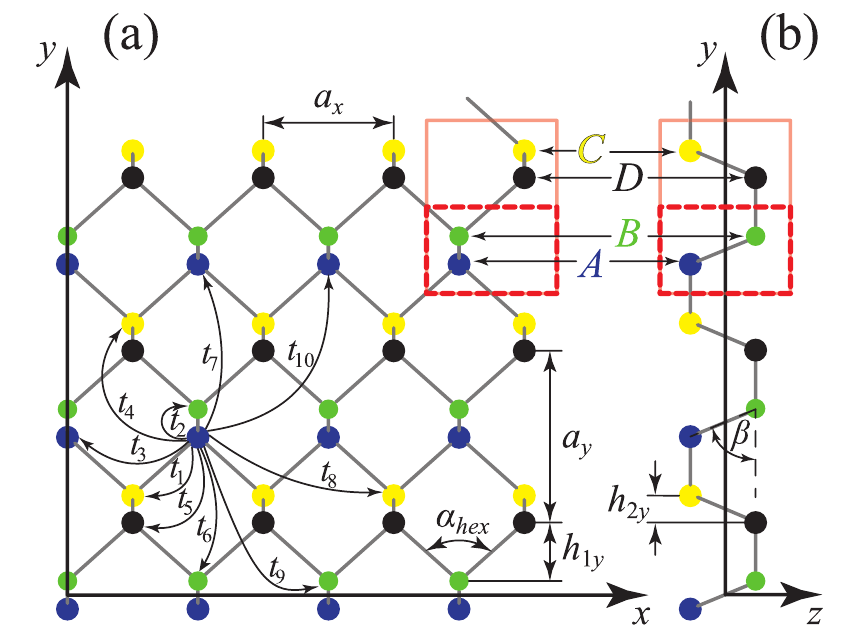}
\caption{Illustration of the phosphorene structure (a) top view, and 
(b) side view. Red solid (dashed) rectangle indicates the unit cell in the four-band 
(two-band) model. Relevant parameters are explicitly indicated in the figure.}
\label{fig1}
\end{figure}

We model the phosphorene nanoribbons using the tight-binding model which is defined by $10$ 
hopping parameters \cite{rudenko15} which are indicated in Fig.~\ref{fig1}. 
Within the tight-binding approximation the Hamiltonian reads
\begin{equation}\label{eq:Hamiltonian}
H=\sum_{i\neq j}t_{ij}e^{i\varphi_{ij}}c^{\dagger}_{i}c_{j},
\end{equation}
where the summation runs over all the lattice sites of phosphorene, $t_{ij}$ 
are the hopping parameters, $\varphi_{ij}=\frac{e}{\hbar}\int_{\mathbf{r_j}}^{\mathbf{r_i}}\mathbf{A}\cdot
d\mathbf{l}$ denotes the Peierls phase picked up while hopping in the presence
of the magnetic field, and $c_j^\dagger (c_i)$ is the creation (annihilation)
operator of an electron on the site {\it j}({\it i}).

The unit cell of the single phosphorene sheet, framed by the solid square line in Fig.~\ref{fig1}, 
contains four atoms labeled by {\it A} (blue), {\it B} (green), {\it C} (yellow) and {\it D} (black).
Using the tight binding model, the four-band Hamiltonian \cite{ezawa14,pereira15}  
for single-layer phosphorene sheet is given by:
\begin{equation}\label{eq:Ham4}
H_{k}^{4\times 4}=\left(\begin{array}{cccc}
	U_{A} & H_{AB} & H_{AD} & H_{AC} \\
	H^*_{AB} & U_{B} & H_{BD} & H_{BC} \\
	H^*_{AD} & H^*_{BD} & U_{D} & H_{DC} \\
	H^*_{AC} & H_{BC}^* & H_{DC}^* & U_{C} \\
\end{array}\right),
\end{equation}
with eigenvectors represented by spinor $[\phi_A \phi_B \phi_D \phi_C]^T$, 
and diagonal terms are given by
\begin{eqnarray}
	U=U_{A-D}=2 t_3 \cos (a_x k_x)+2 t_7 \cos (a_y k_y)\nonumber\\
	+4 t_{10} \cos (a_x k_x) \cos (a_y k_y).
\end{eqnarray}
The interaction terms between sublattice sites are
\begin{align}
	H_{AB}&=H^*_{CD}=t_2 e^{i k_y h_{2y}}+t_6 e^{-i k_y (a_y- h_{2y})}\nonumber\\
	&+2 t_9 \cos (a_x k_x) e^{-i k_y (a_y- h_{2y})},\\
	H_{AD}&=H^*_{BC}=4 t_5 \cos \left(\frac{a_x k_x}{2}\right) \cos \left(\frac{a_y k_y}{2}\right),\\
	H_{AC}&=H^*_{BD}=2 t_4 \cos \left(\frac{a_x k_x}{2}\right) e^{i k_y (a_y- h_{1y})}\nonumber\\
	&+2 \left(t_1\cos \left(\frac{a_x k_x}{2}\right)+t_8\cos \left(\frac{3 a_x k_x}{2}\right)\right)  e^{-i k_y  h_{1y}}.
\end{align}
Here ${h}_{1y}=d_1\cos (\alpha_{hex}/2)$, ${{h}_{2y}}=d_2\cos \beta$ (see Fig.~\ref{fig1}). 
The parameter values are taken from Ref. ~[\onlinecite{rudenko15}].
The equality of certain terms of the Hamiltonian (\ref{eq:Ham4}) indicates the "equivalence" of certain atomic sites. 
This is actually due to the $D_{2h}$ point group invariance \cite{ezawa14, voon15}, which is preserved even when a perpendicular magnetic field is applied. 
Atoms in the upper and lower lattice, namely $A\equiv D$ and $B\equiv C$, are "indistinguishable" and the unit cell is reduced to a single dimmer, 
framed by the dashed square in Fig.~\ref{fig1}. Using this symmetry argument, the 
simplified two-band Hamiltonian reads
\begin{equation}\label{eq:Ham2}
	H_{\mathbf k}^{2\times2}=\left[\begin{array}{cc}U+H_{AD}&H_{AB}+H_{AC}\\
		H^*_{AB}+H^*_{AC}&U+H_{AD}\end{array}\right],
\end{equation}
which acts upon the spinors 
\begin{equation}\label{eq:spinors}
	\psi=\left[\begin{array}{cc}(\phi_A+\phi_D)/2\\
		(\phi_B+\phi_C)/2\end{array}\right]
	=\left[\begin{array}{cc} \phi_1\\ \phi_2\end{array}\right].
\end{equation}
The eigenvalue problem for the Hamiltonian (\ref{eq:Ham2}) can be solved analytically
\begin{eqnarray}
	E=U+H_{AD}\pm|H_{AB}+H_{AC}|,
\end{eqnarray}
where the upper (lower) sign is for the conduction (valence) band. 

Recent theoretical analysis \cite{ezawa14,pereira15} based on the tight binding model with five hopping parameters 
shows that the Maclaurin series of analytical functions in the vicinity of the $\Gamma$-point up to the quadratic term give 
satisfactory accuracy for the phosphorene band structure. Using a similar argument we derive the effective two-band Hamiltonian
\begin{equation}\label{eq:hamiltonian}
	H_{{\mathbf k}\rightarrow\Gamma}^{2\times 2}=\left[\begin{array}{cc}f&g\\
		g^{\dagger}&f\end{array}\right],
\end{equation}
where $f=E_0+\chi_x k_x^2+\chi_y k_y^2$, and
$g=E_g/2-i\gamma k_y+\alpha_x k_x^2+\alpha_y k_y^2$, with
\begin{align}
E_0&=2t_3+4t_5+2t_7+4t_{10}=-0.262 {\rm eV},\\
E_g&=4t_1+2t_2+4t_4+2t_6+4t_8+4t_9=1.838 {\rm eV},\\
\chi_x&=(-t_3-t_5/2-2t_{10})a_x^2=1.060 {\rm eV\AA^2},\\
\chi_y&=(-t_5/2-t_7-2t_{10})a_y^2=-1.772 {\rm eV\AA^2},\\
\alpha_x&=(-t_1/4-t_4/4-9t_8/4-t_9)a_x^2=2.24 {\rm eV\AA^2},\\
\alpha_y&= -(t_1+t_8)h_{1y}^2- t_2/2\cdot h_{2y}^2-t_4\left(a_y-h_{1y}\right)^2\nonumber\\
&-\left(t_6/2+t_9\right)\left(a_y-h_{2y}\right)^2=2.026 {\rm eV\AA^2},\\
\gamma&=2(t_1+t_8) h_{1y}-t_2h_{2y}-2t_4(a_y-h_{1y})\nonumber\\
&+(t_6+2t_9)(a_y-h_{2y})=-5.952 {\rm eV\AA},
\end{align}
and obtain the simplified dispersion for electrons and holes
\begin{eqnarray}\label{eq:dispersion}
E&=&E_0+\chi_xk_x^2+\chi_yk_y^2\nonumber\\
&\pm&\sqrt{\left(\frac{E_g}{2}+\alpha_xk_x^2+\alpha_yk_y^2\right)^2+\gamma^2k_y^2}.
\end{eqnarray}
From Eq.~(\ref{eq:dispersion}) it is obvious that, phosphorene has a non-elliptical dispersion due to the parameter $\gamma$.

In order to transform the low-energy Hamiltonian given by Eq.~(\ref{eq:hamiltonian}) 
into a more convenient form we perform the unitary transformation given by the Hadamard matrix
\begin{equation}
	U=\frac{1}{\sqrt2}\left[\begin{array}{cc}1&1\\
		1&-1\end{array}\right].
\end{equation}
The resulting Hamiltonian has the same form as the one obtained in Ref. ~[\onlinecite{rodin14}] by fitting to 
results from ab-initio calculations 
\begin{equation}\label{eq:hamiltonianU}
		H_{{\mathbf k}\rightarrow\Gamma}^U=\left[\begin{array}{cc}E_{c}&i\gamma k_y\\
		-i\gamma k_y&E_{v}\end{array}\right],
\end{equation}
where diagonal terms
\begin{eqnarray}
	E_c&=&E_0+\frac{E_g}{2}+(\alpha_x+\chi_x)k_x^2+(\alpha_y+\chi_y)k_y^2\nonumber\\
	      &=&E_{c0}+\frac{\hbar^2}{2m_{0e,x}^*}k_x^2+\frac{\hbar^2}{2m_{0e,y}^*}k_y^2,\\
	E_v&=&E_0-\frac{E_g}{2}-(\alpha_x-\chi_x)k_x^2-(\alpha_y-\chi_y)k_y^2\nonumber\\
	      &=&E_{v0}-\frac{\hbar^2}{2m_{0h,x}^*}k_x^2-\frac{\hbar^2}{2m_{0h,y}^*}k_y^2,
\end{eqnarray}
are related to the conduction and valence band dispersions, respectively,  
while the off-diagonal term introduces interband coupling.
Here, $E_{c(v)0}$ are the conduction (valence) band edges in monolayer phophorene, 
$m^*_{0e,x(y)}$ and $m^*_{0h,x(y)}$ are the electron and 
hole effective masses when the coupling is neglected. 
Furthermore, spinors given by Eq.~(\ref{eq:spinors}) are transformed to 
$\psi^U=\left[\phi_c \phi_v\right]^T$, where $\phi_{c(v)}=(\phi_1\pm \phi_2)/\sqrt{2}$.

In the long-wavelength limit $E_c-E_v>E_g\gg|\gamma k_y|$. Thus, to lowest order 
the Eq.~(\ref{eq:dispersion}) is expanded around the $\Gamma$-point and 
the energy dispersions of conduction and valence bands are approximated by 
$E^{eff}_c=E_c+\gamma^2k_y^2/E_g$ and $E^{eff}_v=E_v-\gamma^2k_y^2/E_g$, respectively\cite{zhou15}.
Therefore, the Hamiltonian given by Eq.~(\ref{eq:hamiltonianU}) is reduced to the diagonal form $H_k^{eff,U}\rightarrow {\rm diag}(E^{eff}_c,E^{eff}_v)$ where
\begin{eqnarray}
	E^{eff}_c&=&E_{c0}+\frac{\hbar^2}{2m_{e,x}^*}k_x^2+\frac{\hbar^2}{2m_{e,y}^*}k_y^2,\label{eq:dispersion1}\\
	E^{eff}_v&=&E_{v0}-\frac{\hbar^2}{2m_{h,x}^*}k_x^2-\frac{\hbar^2}{2m_{h,y}^*}k_y^2.\label{eq:dispersion2}
\end{eqnarray}
The effective masses along the $\Gamma-X$ direction remain unchanged, 
\begin{equation}
	m_{e(h),x}^*=m_{0e(0h),x}^*=\frac{\hbar^2}{2(\alpha_x\pm\chi_x)},
\end{equation}
while the effective masses along the $\Gamma-Y$ direction are modified by the term that perturbatively takes into account the interband coupling,
\begin{equation}
	m_{e(h),y}^*=\frac{\hbar^2}{2\left(\alpha_y\pm\chi_y+\frac{\gamma^2}{E_g}\right)}.
\end{equation}
The upper "+" (lower "$-$") sign is for the electron (hole). 
The effective masses along the main axes (see Fig.~\ref{fig1}) are given in Tab. I. 
Note that there is a difference between our results and the one obtained from the previously derived continuum model \cite{zhou15} 
based on 5 hopping parameters\cite{rudenko14}.

Our goal is to analyze ribbons with arbitrary edges, where the translation 
vector of a unit cell is
\begin{equation}
{\mathbf{d}}=c_x a_x{\mathbf{e}}_x+c_y a_y{\mathbf{e}}_y.
\end{equation} 
Here, $c_x$ and $c_y$ are mutually prime integers, ${\mathbf{e}}_{x(y)}$ are the unit vectors, 
while  $d=|{\mathbf{d}} |$ is the length of the unit cell. Since ribbon edges are not flat, 
we define the effective ribbon width as the total square of hexagonal 
plaquettes inside the unit cell divided by the unit cell width $d$.

\begin{table}
\caption{The effective masses along the main axis for the electron and the hole, where $m_0$ is the free electron mass.}
\begin{tabular}{l*{2}{c}}
	\hline
	\hline
		$ $ & $x$-ZZ &{\hspace{1em}} $y$-AC \\
	\hline
	$m_e^*$ (the electron)		& $1.1547m_0$ &{\hspace{1em}} $0.1951m_0$  \\
	$	$				& $0.848m_0^{\clubsuit}$ &{\hspace{1em}} $0.167m_0^{\clubsuit}$  \\
	\hline
	$m_h^*$ (the hole)		& $3.2279m_0$ &{\hspace{1em}} $0.1651m_0$   \\
	$	$				& $1.142m_0^{\clubsuit}$ &{\hspace{1em}} $0.184m_0^{\clubsuit}$   \\
	\hline
	\hline
$^{\clubsuit}$results from Ref.~[\onlinecite{zhou15}]
\end{tabular}
\end{table}

It is straightforward to show that the effective masses of electrons 
and holes in the vicinity of the $\Gamma$-point along  the $x'$ and $y'$ axes, 
which are directed along and perpendicular to the ribbon edge, respectively, can be estimated as:
\begin{eqnarray}
\frac{1}{m_{e(h),x'}^*}=\frac{1}{m_{e(h),x}^*}\cos^2\varphi+\frac{1}{m_{e(h),y}^*}\sin^2\varphi,\label{eq:mex}\\
\frac{1}{m_{e(h),y'}^*}=\frac{1}{m_{e(h),x}^*}\sin^2\varphi+\frac{1}{m_{e(h),y}^*}\cos^2\varphi\label{eq:mey}.
\end{eqnarray}
Here $\varphi=\arctan(c_y a_y/c_x a_x)$ is the angle of rotation with 
respect to the $(x,y)$ coordinate system shown in Fig.~\ref{fig1}. In order to write the Hamiltonian in 
the rotated frame we substitute 
\begin{eqnarray}
	k_x&\rightarrow k_{x'}\cos\varphi-k_{y'}\sin\varphi,\nonumber\\
	k_y&\rightarrow k_{x'}\sin\varphi+k_{y'}\cos\varphi,\nonumber
\end{eqnarray}
in Eqs.~(\ref{eq:dispersion1}) and (\ref{eq:dispersion2}). For infinitely long nanoribbons 
with arbitrary edges the Hamiltonian should not depend on $x'$, therefore $\phi_{c(v)}(x',y')=\phi_{c(v)}(y')\cdot e^{i k_{x'}x'}$. 
For convenience we set $y'=0$ in the middle of the ribbon. 

We include the magnetic field perpendicular to the structure ($ {\mathbf{B}}= B {\mathbf{e}}_z$).
One may show that in the continuum model $k_{x'}\rightarrow k_{x'}-y'/l_B^2$, where we 
adopt the Landau gauge ${\mathbf{A}}=(-By',0,0)$, with $l_B=\sqrt{\hbar/eB}$ being the magnetic field length and operators 
$k_{x'}=-i\partial/ \partial x'$ and $k_{y'}=-i\partial/ \partial y'$. 
Using Eqs.~(\ref{eq:mex},\ref{eq:mey}) and after some elaborate algebra we obtain a set of 
decoupled differential equations for the conduction and valence band,
\begin{eqnarray}
	\frac{\hbar^2}{2m^*_{e(h),y'}}\left[-i\frac{\partial}{\partial\tilde{y}}-\tilde{y} \left(\frac{m^*_{e(h),y'}}{m^*_{e(h),x'}}-1\right) \frac{\tan2\varphi}{2l^2_B}\right]^2 \nonumber\\
+\frac{1}{2}m^*_{e(h),y'}\omega_{e(h)}^2\tilde{y}^2 = \Delta E_{c(v)},\label{eq:masterD}
\end{eqnarray}
where the subscripts $c$ and $e$ are used for the conduction band, and $v$ and $h$ denote the valence band.
Here, $\tilde{y}=y'-k'_x l^2_B$ is the shifted $y'$ coordinate, 
$\omega_{e(h)}=eB/\sqrt{m^*_{e(h),x}m^*_{e(h),y}}$ are the cyclotron frequencies for the 
electron (hole) in bulk phosphorene, and $\Delta E_{c(v)}=\pm (E_{c(v)}-E_{c(v)0})$ are 
eigenenergies measured from the bottom (top) of the conduction (valence) band edges. 
One may notice that the above equation does not include the term related to the 
band offset at the edges of the ribbon. This is because we want to analyze the conditions 
under which LLs exist in a ribbon. Namely, we insist that, either the ribbon is sufficiently wide, 
or the magnetic field is sufficiently strong, so that particle localization is governed by the effective parabolic potential 
$V_{e(h)}(k_{x'},y',B)=\frac{1}{2}m_{e(h),y'}\omega_{e(h)}^2\left(y'-l^2_B k_x'\right)^2$ 
given in Eq.~(\ref{eq:masterD}), rather than by the ribbon edges.

We seek the eigenvector components in the form
\begin{eqnarray}
	\phi_{c(v)}(\tilde{y})=\exp\left(-\frac{m^*_{e(h),y'}\omega_{e(h)}}{\hbar}\frac{\tilde{y}^2}{2}\right)\nonumber\\
	\times \exp\left[ i \left( \frac{m^*_{e(h),y'}}{m^*_{e(h),x'}}-1\right)\frac{\tan2\varphi}{2l^2_B} \frac{\tilde{y}^2}{2}\right]f_{c(v)}(\tilde{y}),\label{eq:wf}
\end{eqnarray}
which leads to the differential equation
\begin{equation}
	f_{c(v)}''(\xi)-2\xi f'_{c(h)}(\xi)+(\varepsilon_{c(v)}-1)f_{c(v)}(\xi)=0,\label{eq:wff}
\end{equation}
where $\xi=\xi_{e(h)}=\sqrt{m^*_{e(h),y'}\cdot\omega_{e(h)}/\hbar}\cdot \tilde{y}$ is the dimensionless coordinate and 
$\varepsilon_{e(h)}=2\Delta E_{c(v)}/\hbar\omega_{e(h)}$ is the dimensionless energy.
Finally, to get eigenvalues denoted by the integer quantum number we impose the condition $\varepsilon_{c(v)}-1=2n_{e(h)}$. 
Thus, the solutions of Eq.~(\ref{eq:wff}) are Hermite polynomials
\begin{equation}
	f_{c(v)}(\xi)=C_n H_{n}(\xi)=C_n (-1)^n e^{\xi^2}\frac{d^ne^{-\xi^2}}{d\xi^n},\label{eq:sf}
\end{equation}
where the principle quantum number $n\triangleq n_{e(h)}=\{0,1,2,...\}$ is the LL number, 
 the normalization constant is $C_n=\sqrt[4]{m^*_{e(h),y'}\cdot\omega_{e(h)}}/\sqrt{{n!}2^n\pi^{1/2}\hbar^{1/2}}$,
and the eigenvalues of the LLs follow the quantization of a one-dimensional quantum harmonic oscillator (QHO)
\begin{eqnarray}\label{eq:dispersions}
	E^{LL}_c&=&E_{c0}+\hbar\omega_e\left(n_e+\frac{1}{2}\right),\label{eq:dispersionLHO1}\\
	E^{LL}_v&=&E_{v0}-\hbar\omega_h\left(n_h+\frac{1}{2}\right).\label{eq:dispersionLHO2}
\end{eqnarray}

It is obvious that the separation between adjacent energy levels ($|\Delta n_{e(h)}|=1$) is approximately 
$\Delta E_{c(v)}=\hbar\omega_{e(h)}$. This energy difference is the same for ribbons independent 
of edges and is equal to the one found in single-layer phosphorous sheet\cite{zhou15,pereira15}.

It should be pointed out that the exact treatment of LLs in bulk phosphorene shows that 
the off-diagonal terms, that accounts for the conduction and valence band coupling, 
are similar in form as the Rasshba (Dresselhaus) spin-orbit interaction in conventional semiconductors\cite{winkler03}. 
However, these terms do not contribute significantly to the spectra of the lowest energy states 
and the spatial density distribution corresponding to the first few LLs are found to have elliptical shape\cite{zhou15}. 
In a quasi-classical picture we might infer that, when the magnetic field is turned on perpendicular to the bulk 
phosphorene sheet, electrons and holes undergo elliptical cyclotron orbits.

Let us now consider the impact of a perpendicular magnetic field on the formation of LLs. 
When the magnetic field is turned on, the particle is essentially confined in the transversal 
direction by a restricted effective parabolic potential $V_{e(h)}(k_{x'},y',B)$. 
One may notice that on increase (decrease) of $k_{x'}$ shifts the effective parabolic potential towards 
the upper (lower) ribbon edge, for both the electron and the hole. Therefore, states occupying positive 
and negative momenta reside on opposite sides of the ribbon. Also, the magnetic field does not split
oppositely charged particles in the transversal direction, and the electron and the hole will have 
similar localization in space even when the magnetic field is turned on. Furthermore, the effective potential is 
proportional to the transverse effective mass $m^*_{e(h),y'}$. Thus, the confinement will be 
stronger in ribbons with higher effective mass in the transversal direction.

For relatively low values of $B$, the influence of the parabolic potential is small and the wave function is essentially 
determined by edges of the ribbon, namely, the potential in transversal direction resembles an infinite potential well. 
However, when eigenvalue energy of the $n_{e(h)}$-th electron (hole) state in bulk phosphorene is smaller (larger)
than the effective potential at the ribbon edge closer to the potential extrema, i.e. when
\begin{equation}\label{eq:condition}
	\hbar\omega_{e(h)}\left(n_{e(h)}+\frac{1}{2}\right)
	< \frac{1}{2}m_{e(h),y'}\omega_{e(h)}^2\left(\frac{W}{2}-l^2_B| k_{x'}|\right)^2,
\end{equation}
the confinement along the $y'$ direction is dominantly QHO-like. We should note that this 
criteria is not so rigid. Namely, the effective potential at the closer edge should be 
sufficiently larger than the QHO eigenenergy so that the corresponding eigenfunction decreases sufficiently 
before it reaches the ribbon edge. More comprehensive criteria are established in Ref. ~[\onlinecite{topalovic16}], 
but since we have already made a few approximations it would not improve the analytical results significantly.

The smallest value of the magnetic field for LL formation in the $n$-th electron(hole) state is found 
when equality is replaced by $<$ in Eq.~(\ref{eq:condition}) and $k_{x'}=0$ 
\begin{equation}\label{eq:Bmin}
	B_{n_{e(h)}}^{\min}=\frac{4\hbar(2n_{e(h)}+1)}{eW^2}\frac{\sqrt{m^*_{e(h),x}m^*_{e(h),y}}}{m^*_{e(h),y'}}.
\end{equation}
Note that the minimal field $B_{n_{e(h)}}^{\min}$ is proportional to the number of the LL.
In the above equation only the transverse effective mass $m^*_{e(h),y'}$ depends 
on the edge orientation, and $B_{n_{e(h)}}^{\min}$ monotonically decreases with it. 
In phosphorene the largest value of the effective mass is in the zigzag direction, 
while the smallest one is in the armchair direction. Therefore, the magnetic field 
required for the formation of the LLs is the smallest in the case of AC ribbons and the 
largest for ZZ ribbons. Similar conclusion could be drawn intuitively from the analysis of the 
effective potential, based on the argument related to the confinement strength.

Furthermore, when $B>B^{\min}_{n_{e(h)}}$ we deduce from Eq.~(\ref{eq:condition}) that in the range 
$k_{x'}\in\left(-k_{n_{e(h)},x'}^{FB},k_{n_{e(h)},x'}^{FB}\right)$ 
eigenvalues become independent on $k_{x'}$ where
\begin{eqnarray}
	k_{n_{e(h)},x'}^{FB}=\frac{1}{l_B}\Bigg[\frac{W}{2l_B}-\sqrt{2n_{e(h)}+1}\nonumber\\
	\times \left(\frac{\sqrt{m^*_{e(h),x}m^*_{e(h),y}}}{m^*_{e(h),y'}}\right)^\frac{1}{2}\Bigg],
\end{eqnarray} 
is the {\it flat-band boundary wavenumber}.
Namely, the band structure of states which satisfy the condition (\ref{eq:condition}) 
should appear flat. It is straightforward to 
show that the flat-band boundary wavenumber increases with transverse mass. 
Therefore, the flat-band range is the widest for a ribbon with AC edges and the smallest for ZZ ribbon.
Furthermore, the width of the flat-band decreases with increasing LL number, 
which is due to the fact that higher states extend more along the width of the ribbon, so the 
condition that the wavefunction "touches" the ribbon edge given by Eq.~(\ref{eq:condition}) 
is satisfied for smaller values of the longitudinal momentum.

Let us also briefly discuss the influence of an in-plane electric field $E_{y'}$ on the LLs. 
Such an external electric field modifies the on-site energy. Therefore, in the continuum model we add the potential
$V^{ext}(E_y',y')=eE_{y'}y'=eE_{y'}(\tilde{y}+l^2_Bk_{x'})$ to the diagonal terms. 
It is straightforward to show that the differential equations (\ref{eq:masterD}) are modified so that 
the effective potential becomes
$V_{e(h)}(k_{x'},y',B)=\frac{1}{2}m_{e(h),y'}\omega_{e(h)}^2\left(y'-l^2_B k_x'\pm eE_{y'}/(m_{e(h),y'}\omega^2_{e(h)})\right)^2$ 
while two terms are added to the right-side of the equations 
$\Delta E_{c(v)}\rightarrow \Delta E_{c(v)}+eE_{y'}k_{x'}l^2_B\mp e^2 E^2_{y'}/(2m_{e(h),y'}\omega^2_{e(h)})$. 
Finally, we obtain similar solutions for the eigenfunctions as for the case when only the magnetic field is applied. 
As expected, the wavefunctions are shifted along the ribbons width in the direction of transverse electric 
field for holes, and in the opposite direction for electrons. Also, eigenenergies of LLs are modified 
\begin{equation}\label{eq:LLmod}
	E_{c(v)}=E^{LL}_{c(v)}(E_{y'}=0)\mp\frac{e^2E^2_{y'}}{2m_{e(h),y'}\omega^2_{e(h)}}+eE_{y'}k_{x'}l^2_B.
\end{equation}
We infer that energy gap, between the states in the conduction and valence bands with the same $k_{x'}$, 
is reduced by $e^2E^2_{y'}/(m_{e(h),y'}\omega^2_{e(h)})$, while formerly flat bands adopt linear 
dispersion ($\mathtt{\sim} eE_{y'}k_{x'}l^2_B$) turning the band-gap to indirect. This behavior is expected 
since $B$ leads to a shift of states with positive (negative) momenta to the upper (lower) side of the ribbon, and therefore 
electrons and holes experience opposite potential shifts. Moreover, the wavenumbers 
that determine the boundaries of these linear regions can be found from Eq.~(\ref{eq:condition}) 
when $k_{x'}$ is substituted by $k_{x'}\mp \Delta k(E_{y'},B)$, where 
$\Delta k(E_{y'},B)=eE_{y'}/(m_{e(h),y'}\omega^2_{e(h)}l^2_B)=(E_{y'}/B\hbar)\cdot m_{e(h),x} m_{e(h),y}/m_{e(h),y'}$. 
One may note that these spectral shifts depend linearly on $E_{y'}$. Moreover, based on the argument 
regarding the value of the effective mass in a certain direction, it is easy to conclude that 
these shifts are smallest in a ribbon with AC edges and are largest for the one with ZZ edges. 
Consequently, the band dispersions become tilted when $E_{y'}$ is applied.

The presented continuous model does not account for the edge states. 
The presence and the origin of these states in ribbons with various edges were discussed 
in detail in Refs.~[\onlinecite{grujic16},\onlinecite{ezawa14}]. 
Recently proposed continuous model with proper boundary conditions results in adequately modeled 
edge states for ZZ ribbons\cite{sousa16}. In former approach \cite{ezawa14} the edge states are treated as 
quasi-flat bands and are determined as the zero-energy states in the anisotropic honeycomb lattice model.

We note that for chosen directions of the magnetic and electric fields, the wavefunctions 
are symmetric along $x'$ direction of the ribbon's unit cell. Therefore, the effective translation 
vector that is twice shorter $({\mathbf d}/2)$ can be introduced. The same argument is used for 
the derivation of the two-band Hamiltonian. As a consequence, the First Brillouin zone (FBZ) becomes 
twice wider and the energy spectrum unfolds. The calculation times become more that twice shorter.

\section{Nanoribbons in external perpendicular magnetic and transverse electric field}

\begin{figure} \centering
\includegraphics[width=8.6cm]{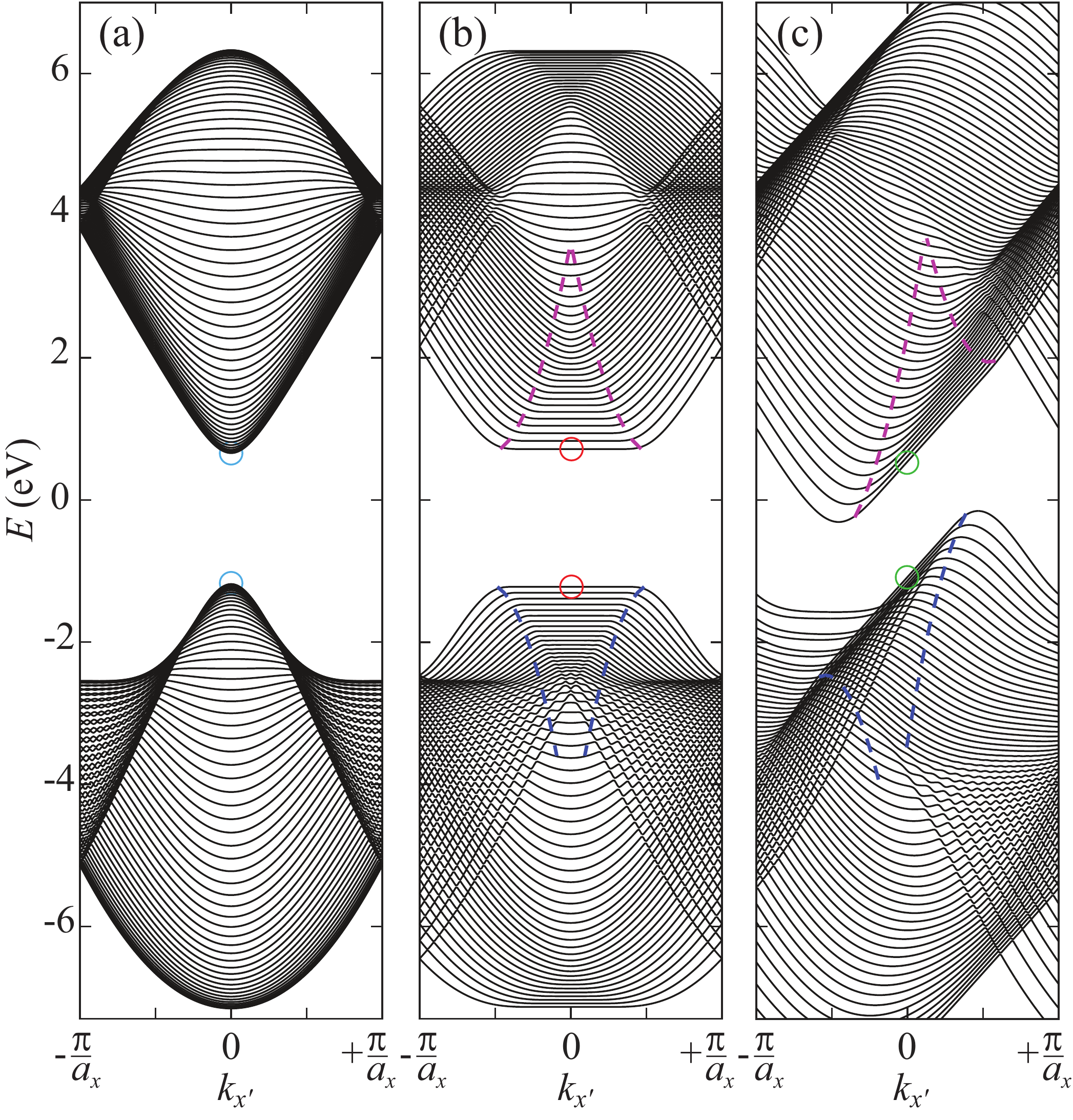}
	\caption{Bend structure of the $N_{AC}=61$ nanoribbon for (a) $B=0$ and $E_{y'}=0$,
	(b) $B=500$ T and $E_{y'}=0$, and (c) $B=500$ T and $E_{y'}=20$ mV/\AA. 
	Dashed magenta (blue) lines in the conduction (valence) bend denote theoretically calculated edges of the 
	flat and linear bands in (b) and (c), respectively.}
\label{fig2}
\end{figure}

\begin{figure} \centering
\includegraphics[width=8.6cm]{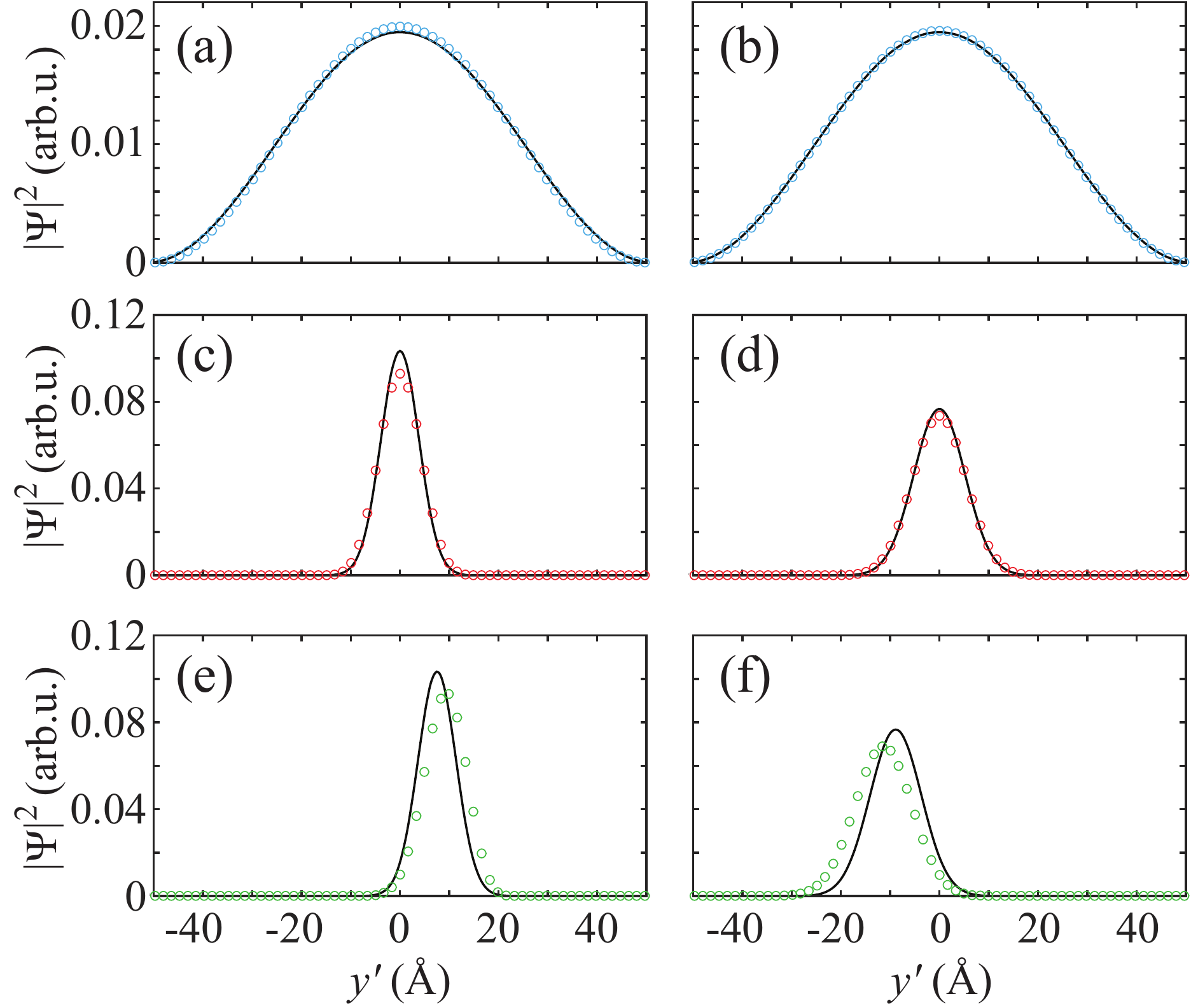}
	\caption{Left (right) column depicts the probability density of the states in the $N_{AC}=61$ nanoribbon
	in the top of the valence (bottom of the conduction) band at $k_{x'}=0$ when (a,b) $B=0$ and $E_{y'}=0$ (first row),
	(c,d) $B=500$ T and $E_{y'}=0$ (second row), and (e,f) $B=500$ T and $E_{y'}=10$ mV/\AA (third row). The solid curves
	and the symbols show $|\Psi|^2$ found from the continuum and the tight-binding model, respectively.}
\label{fig3}
\end{figure}

Our goal is to compare nanoribbons with equal width having different edges. Therefore, we 
chose the number of dimmers along the ribbon cross-section to be 61, 38, 46 and 74, so we have approximately 10 nm wide 
nanoribbons with AC, sZZ, ZZ and sAC edges, respectively. In order to establish 
conditions for LL formation in all nanoribbons, we chose the value of perpendicular 
applied magnetic field $B=500$ T which is rather large in order to have significant spacing between LL's 
which improves its visualisation. Similar results are obtained for smaller $B$ if we increase the width $W$. 

In Fig.~\ref{fig2} we show the band structures of AC nanoribbons:
(a) in the absence of fields, (b) with applied magnetic field only, and (c) with both $B$ and $E_{y'}$ turned on. 
In the absence of fields these nanoribbons are insulating, \cite{grujic16} and host a series of parabolic bands, which are
effectively "sampled" from the band structure of bulk phosphorene, as might be inferred from Fig.~\ref{fig2}(a). 
When the magnetic field is turned on LLs are formed and flat-bands appear, as shown in Fig.~\ref{fig2}(b). Quasiclassically,
for states away from the edges the magnetic field enforces closed elliptical cyclotron orbits. These 
states in turn quantize into LLs and form the flat parts of the bands displayed in Fig.~\ref{fig2}(b). 
Note that these segments get narrower at higher energies and lower magnetic fields; the reason 
being that the cyclotron radius enlarges, and therefore fewer orbits are uninterrupted by the edges.
 
As we discussed in the theory part, due to the large effective mass in the transversal ZZ direction confinement 
is strong, and LLs might be found almost as soon as $B$ is turned on. Therefore, a large number of LLs is 
supported by nanoribbons with AC edges. Dashed magenta and blue curves denote theoretically calculated edges 
of the flat-bands in the conduction and valence band, respectively. There is good agreement between these edges and our numerical 
results. One of the reasons for certain small differences is explained in Sec. II, and is related to the 
flexible condition for these edges, given in Eq.~(\ref{eq:condition}). The other reasons are finite width 
of the ribbon and interband coupling, \cite{zhou15} due to which the spacing between LLs decreases as we 
move from the top (bottom) of the valence (conduction) band. Therefore, $\Delta E^{LL}_{c(v)}$ are 
overestimated and slopes of the theoretically calculated edges are higher than the actual ones. Furthermore, 
we must bear in mind that the continuum approximation is valid in a relatively narrow range of momenta around the 
$\Gamma$-point. Consequently, the discrepancy between the analytical and numerical results is larger for the AC ribbon.

When an electric field $E_{y'}$ is applied the flat bands become linearly dependent on $B$ and appear as tilted. 
As predicted by theory, due to the linear term the bandgap switches to an indirect one, as soon as the in-plane 
electric field is applied. For the value of the electric field $E_{y'}=20 {\rm {mV/\AA}}$ the band-gap closes, 
as might be observed from Fig.~\ref{fig2}(c). Furthermore, shifts of these linear segments in momentum 
space seem to be almost equal for the conduction and valence band. Earlier, we analyzed the spectral shift with respect to the effective 
masses and found that $\Delta k\sim m_{e(h),x}m_{e(h),y}/m_{e(h),y'}$. In the case of AC ribbons 
$\Delta k^{AC}\sim m_{e(h),y}$, where $m_{e,y}=0.195m_0$ and $m_{h,y}=0.165m_0$, thus the spectral shifts 
have similar values for electrons and holes. We note that only for AC ribbons the shift is slightly larger for electrons than for holes. 
For all the other ribbons discussed in the paper, the mass-dependent part of the shift for the holes 
is approximately 1.5-3 times larger than for the electrons. Therefore, we expect that for ribbons other than AC, 
the holes are much more sensitive to the in-plane field than the electrons.

In Fig.~\ref{fig3} we show the real-space probability density of the states at the top (bottom) of the 
valance (conduction) band for $k_{x'}=0$. 
The states are marked with correspondingly colored circles in Fig.~\ref{fig2}. Numerical results 
are compared to the analytical solutions. In order to compare probability densities found by the tight-binding 
model to those determined by means of the continuum approximation, we divide the probability 
for occupying a certain state by the distance between the adjacent atoms in each sublattice that is $\Delta y'_{AC}=a_r/2$ for AC and 
$\Delta y'=\frac{c_r}{2}\cos\varphi=\frac{|a_t|}{d}\frac{a_rc_r}{2}$ for all the other. When magnetic and 
electric fields are not present, confinement along transversal direction is as in an infinite potential well. 
Wavefunctions that are obtained numerically (denoted by open circles in Fig.~\ref{fig3}) are compared to the theoretical expression 
\begin{equation}
	\Psi_0(y',k_{x'}=0)=\sqrt{\frac{2}{W}}\cos\left(\frac{\pi}{W} y'\right)\nonumber
\end{equation}
displayed by solid lines in Figs.~\ref{fig3}(a) and (b), for the hole and the 
electron, respectively. We found that the wavefunction is spread along the ribbons with similar distribution for both, the electron 
and the hole, as predicted by the theoretical expression. According to expression (\ref{eq:wf}) for the 
case when both magnetic and electric fields are applied, the ground state wavefunction at $k_{x'}=0$ reads 
\begin{eqnarray}
	&|\psi^{QHO}_{0,e(h)}(y',k_{x'}=0)|=\left(\frac{m_{e(h),y'}\omega_{e(h)}}{\pi \hbar}\right)^\frac{1}{4}\nonumber\\
	&\times \exp\left[-\frac{m_{e(h),y'}\omega_{e(h)}}{2 \hbar}\left(y'\pm\frac{eE_{y'}}{m_{e(h),y'}\omega^2_{e(h)}}\right)^2\right]. 
\end{eqnarray}
When only the magnetic field is present we found excellent agreement between our numerical 
results and the theoretically predicted Gaussian function from our simplified model. Namely, almost perfect bell-shaped functions 
are formed in the middle of the ribbon, as may be inferred from Figs.~\ref{fig3}(c) and (d) for the hole 
and electron, respectively. Good agreement is also found in the presence of electric field as shown in Figs.~\ref{fig3}(d) and (e). 
The wavefunctions remain Gaussian-like but shift toward upper (lower) edge for the hole (electron).

\begin{figure} \centering
	\includegraphics[width=8.6cm]{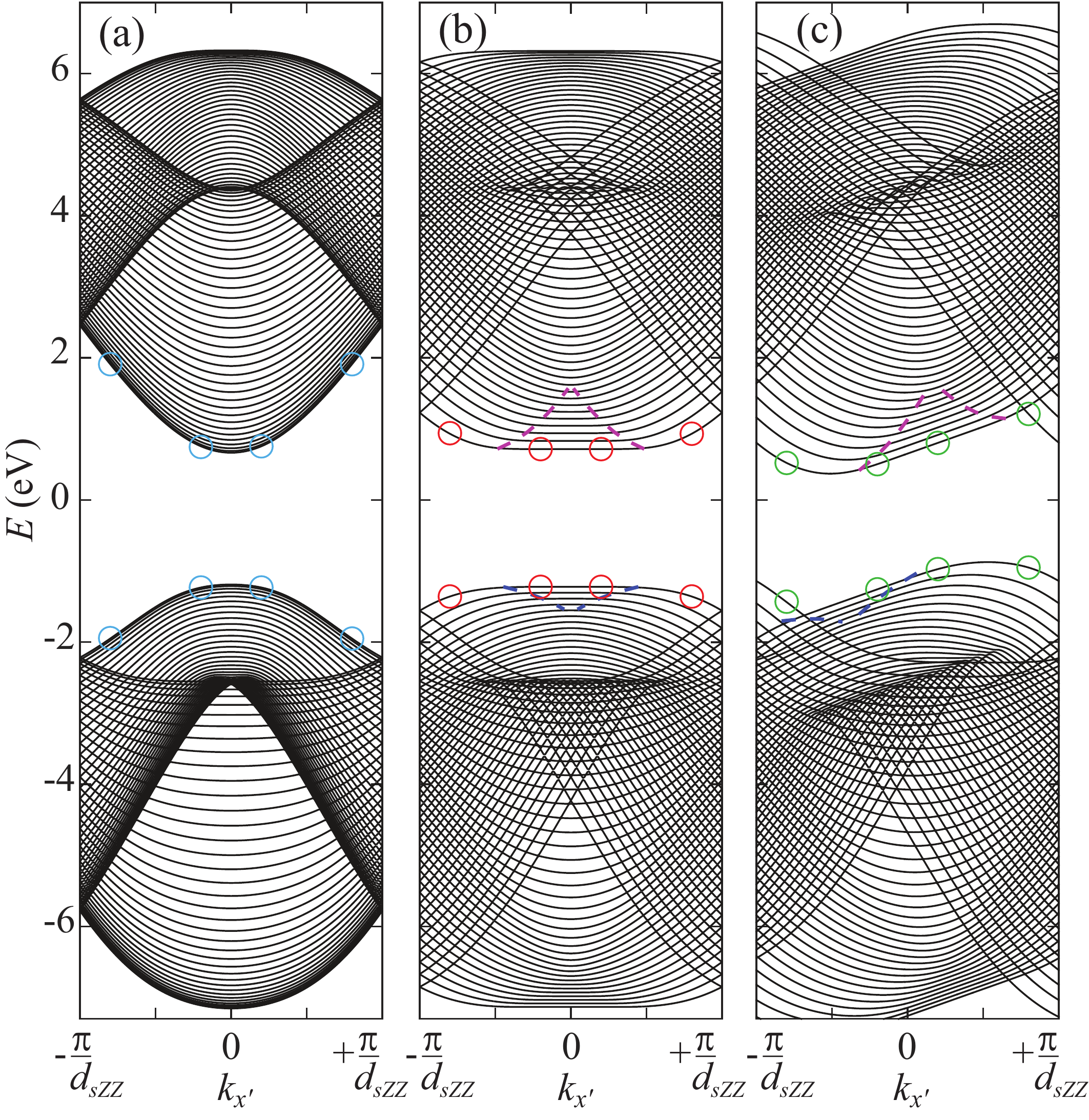}
	\caption{The band structure of $N_{sZZ}=38$ nanoribbon for (a) $B=0$, and $E_{y'}=0$,
	(b) $B=500$ T and $E_{y'}=0$, and (c) $B=500$ T and $E_{y'}=10$ mV/\AA.}
	\label{fig4}
\end{figure}

\begin{figure} \centering
	\includegraphics[width=8.6cm]{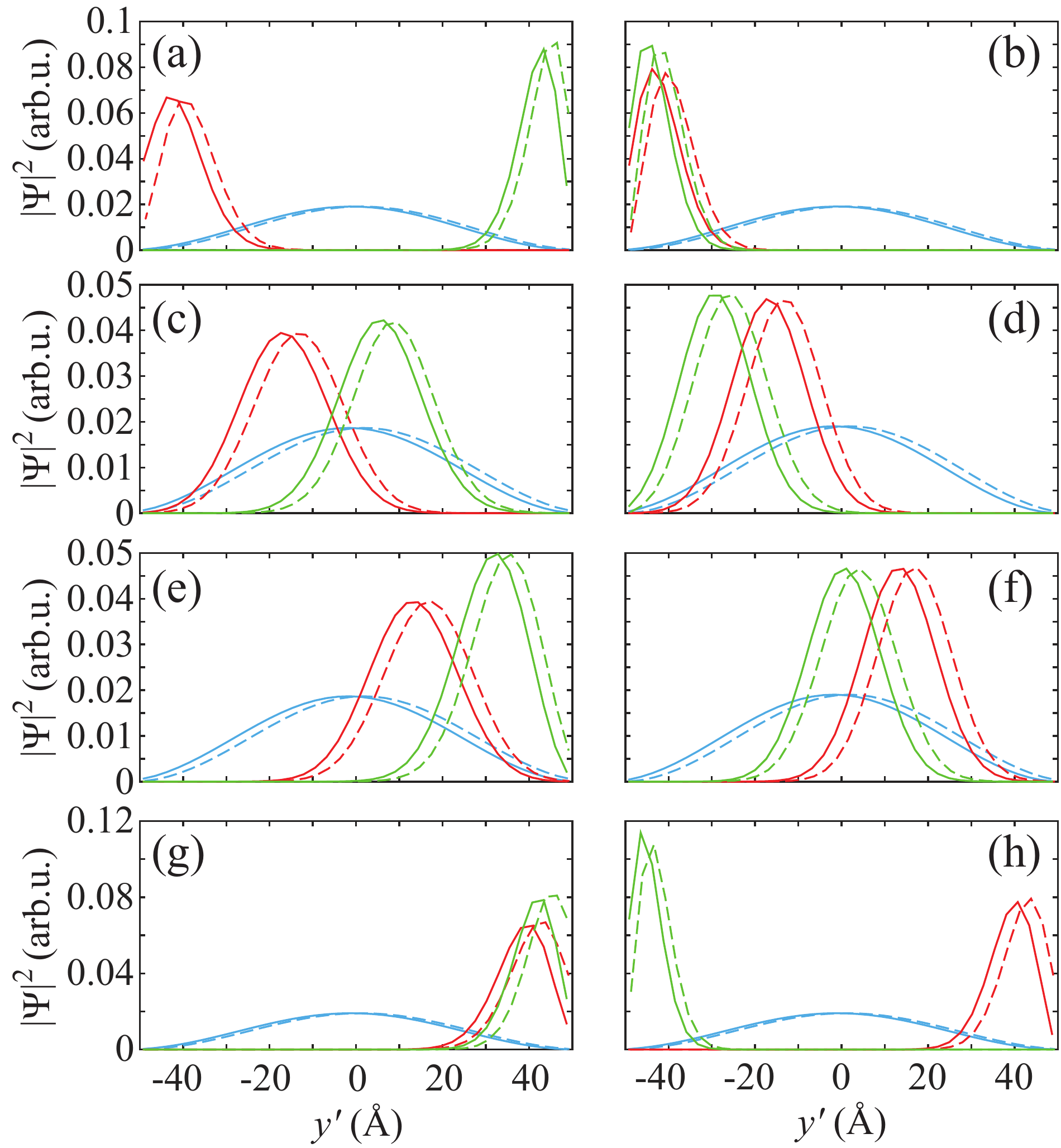}
	\caption{The left (right) column depicts the probability density of the ground states of the $N_{sZZ}=38$ nanoribbon
	in the top of the valence (bottom of the conduction) band at $k_{x'}=-0.8\pi/d_{sZZ}$ (first row), 
	$k_{x'}=-0.2\pi/d_{sZZ}$ (second row), $k_{x'}=0.2\pi/d_{sZZ}$ (third row) and $k_{x'}=0.8\pi/d_{sZZ}$ (fourth row). 
	The light blue lines correspond to $B=0$ and $E_{y'}=0$, red lines correspond to $B=500$ T and $E_{y'}=0$, 
	while green lines correspond to $B=500$ T and $E_{y'}=10$ mV/\AA. The solid and dashed lines show $|\Psi|^2$
	of the opposite sublattices.}
	\label{fig5}
\end{figure}

Next, we display the evolution of the band structure of sZZ nanoribbons in magnetic and electric fields 
in Fig.~\ref{fig4}. In the absence of the fields, similarly for AC, these ribbons are insulating, as inferred from Fig.~\ref{fig4}(a). 
When sufficiently large $B$ is applied, LLs are formed in the middle of the FBZ, i.e. the parabolic bands evolve, and start 
featuring dispersionless segments as we observed earlier in Fig.~\ref{fig4}(b). By inspection of Fig.~\ref{fig4}(b) we note that 
there is a smaller number of flat segments that are somewhat narrower than in the case of the AC ribbon with 
approximately the same width.

In order to elucidate these effects, in Fig.~\ref{fig5} we plot the real-space probability density 
of the ground states marked with correspondingly colored circles in Fig.~\ref{fig4}, where solid and dashed curves 
depict $|\Psi|^2$ on opposite sublattices. Both sublattices exhibit the same functional variation of 
$|\Psi|^2$, with a slight lateral offset, which is a general property, independent of the applied fields. 
The left (right) panel column corresponds to the valence (conduction) band, while the first, second, third 
and fourth row in the panel correspond to $k_{x'}=-0.8\pi/d$, $k_{x'}=-0.2\pi/d$, $k_{x'}=0.2\pi/d$, and $k_{x'}=0.8\pi/d$, respectively. 
In the absence of fields the ground electron and hole states fully extend across the ribbon width, regardless of the 
$k_{x'}$, as might be observed from the light blue curves in Figs.~\ref{fig5}(a-h).  For nonzero $k_{x'}$ energy increases as described 
by $\hbar^2k^2_{x'}/2m_{e(h),x'}$, while confinement along ribbon width remains unchanged.  
The probability density of the ground states corresponding to the dispersionless segments in the valence (conduction) 
band are shown by the red curves in Figs.~\ref{fig5}(c-f). We infer that these probability densities have almost unchangeable 
Gaussian shape for any $k_{x'}$ in the flat-band range, as can be concluded by comparing the red curves in 
Figs.~\ref{fig5}(c) and (e), for the valence band, as well as from Figs.~\ref{fig5}(d) and (f), for the conduction band. 

\begin{figure} \centering
	\includegraphics[width=8.6cm]{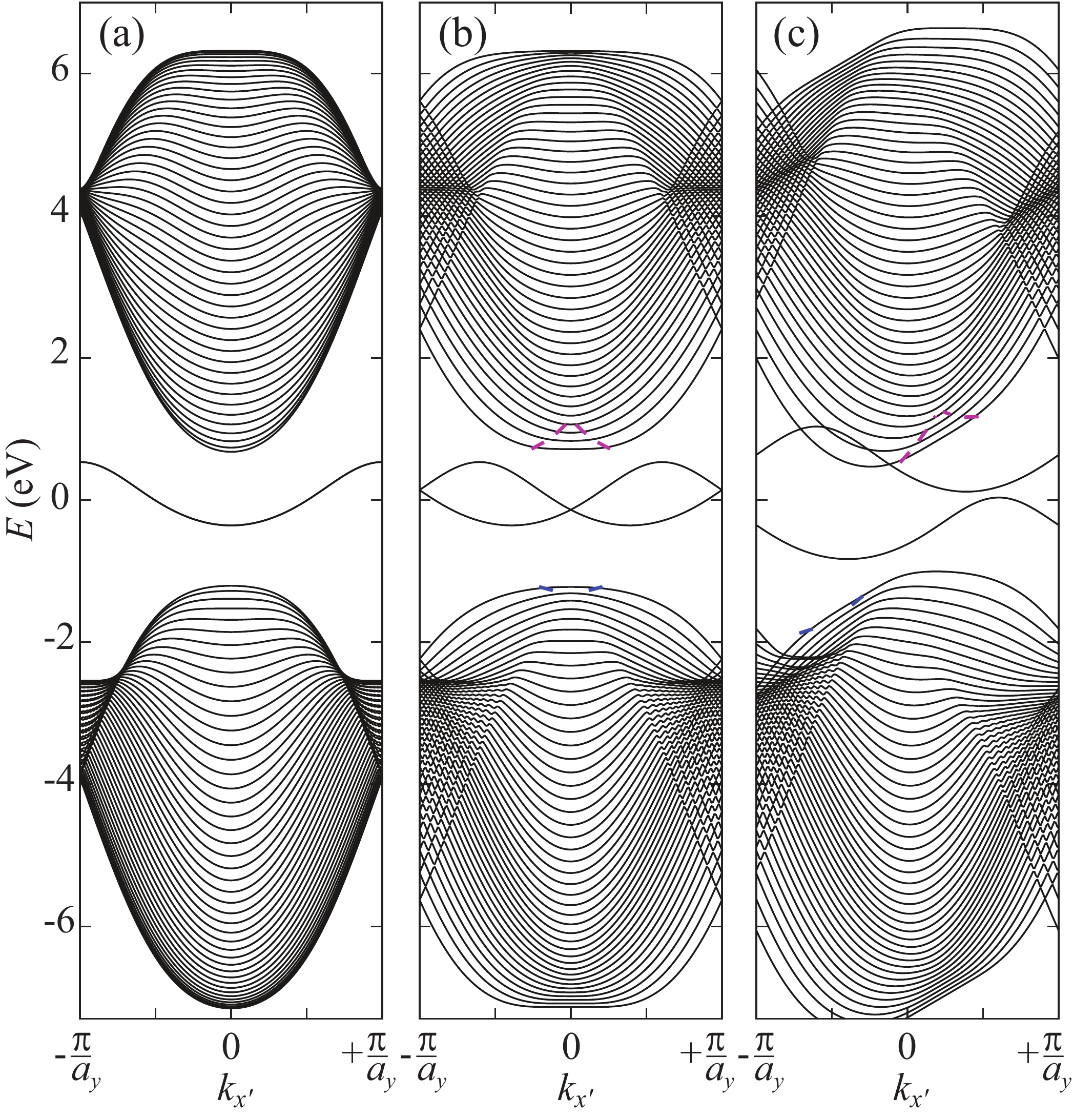}
	\caption{The band structure of $N_{ZZ}=46$ nanoribbon for (a) $B=0$, and $E_{y'}=0$,
	(b) $B=500$ T and $E_{y'}=0$, and (c) $B=500$ T and $E_{y'}=10$ mV/\AA.}
	\label{fig6}
\end{figure}

Next, we analyze effects of applied transverse electric field. By comparing the green curves in Figs.~\ref{fig5}(b-g) to the red ones 
we found that the states are almost the same as in the case when only the magnetic field is applied. As expected, 
calculated wavefunctions in linear segments of the spectrum shown in Fig.~\ref{fig4}(c) have the shape that is almost identical to 
the corresponding eigenfunctions in the flat-bands. However, these states are shifted along the ribbon width due 
to the electric field. For negative values of the longitudinal momenta where the dispersion is linear 
(see linear segments limited by dashed blue line in Fig.~\ref{fig4}(c)), the hole wavefunction is 
localized around the center of the ribbon.  Note that the valence band ground state at $k_{x'}=0.2\pi/d$ is not in the linear region, as 
shown in Fig.~\ref{fig4}(c). In fact, since $\Delta k^{sZZ}_h/\Delta k^{sZZ}_e=2.352$ linear segments are much more shifted in 
the valence than in the conduction band, as displayed in Fig.~\ref{fig4}(c).   Therefore, the probability density displayed in Fig.~\ref{fig5}(e) interferes with 
the upper ribbon edge, and differs from the bell-shaped function shown in  Fig.~\ref{fig5}(c).

However, the most intriguing behavior occurs near the left zone edge, where crossing is found for the hole ground state. 
The crossing involves the states localized at the opposite edges of the ribbon (see Fig.~\ref{fig4}(c)). Therefore, 
the localization of the hole ground state can be abruptly changed by the electric field, as shown in Fig.~\ref{fig5}(a). 
The same effect occurs in the conduction band, but near the upper edge of the FBZ (see Fig.~\ref{fig4}(c)). 
Due to the crossing the electron localization switches to the opposite edge, as shown by the green curves in Fig.~\ref{fig5}(h). 
Additionally, the position of the crossings can be tuned by magnetic field.

\begin{figure} \centering
	\includegraphics[width=8.6cm]{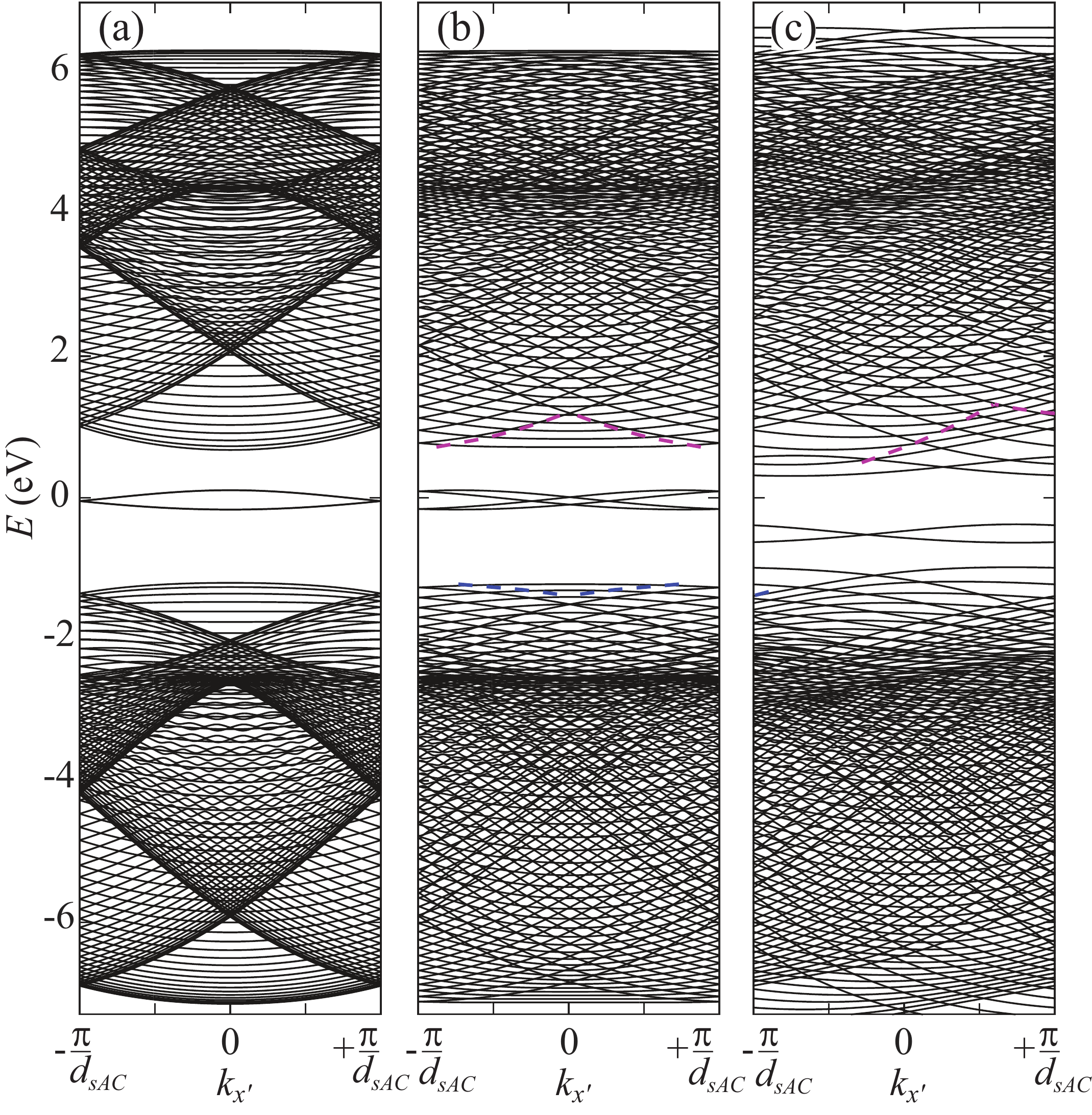}
	\caption{The band structure of $N_{sAC}=74$ nanoribbon for (a) $B=0$, and $E_{y'}=0$,
	(b) $B=500$ T and $E_{y'}=0$, and (c) $B=500$ T and $E_{y'}=10$ mV/\AA.}
	\label{fig7}
\end{figure}

The band structures of zigzag nanoribbons are shown in Fig.~\ref{fig6}. As discussed in the 
theoretical part, the interband coupling has a smaller influence on the motion along the $\Gamma$-X direction 
and the dispersion is almost perfectly parabolic in most of the bands (see Fig.~\ref{fig6}(a)). 
Since ZZ ribbons have weakest confinement in transversal direction even when a high $B$ is 
applied, only a few LLs evolve (see small regions encircled by dashed lines in Fig.~\ref{fig6}(b)).
When the electric field is applied, bands become immediately tilted, and newly formed linear bands are 
shifted. By comparing these shifts in the conduction and valence bands in Fig.~\ref{fig6}(c), we realize 
that the hole is much more sensitive to the electric field, which is supported by the fact that the ratio  
$\Delta k^{ZZ}_h/\Delta k^{ZZ}_e=2.795$ for ZZ nanoribbons is the highest out of all considered ribbons.

In Fig.~\ref{fig7}, we show the band structure of skewed armchair nanoribbons. Note that there are less flat segments in the spectrum of 
Fig.~\ref{fig7}(b) corresponding to LLs forming around the middle of the ribbon than in the case of AC 
and sZZ ribbons. Also, these segments are much narrower, which can be explained by having in mind 
that the width of FBZ for sAC is 1.5-3 times narrower than for the other ribbon types. 
Since $\Delta k^{sAC}_h/\Delta k^{sAC}_e=2.734$ linear segments are much more shifted in 
the valence than in the conduction band, and dashed blue line delimiting linear bands starts at the 
band edge, as shown in Fig.~\ref{fig7}(c). 

Finally, we investigate the behavior of edge states. These states do not undergo Landau quantization, 
as can be seen in Fig.~\ref{fig6}(b) and Fig.~\ref{fig7}(b), which is a consequence of their exponential 
localization near the edges.\cite{grujic16} Instead, each QFB is effectively shifted along the $k_{x'}$ axis, 
but in opposite directions. This behavior was absent in the case of edge states in zigzag graphene 
ribbons, since the corresponding bands were not fully detached from the bulk bands,\cite{ezawa14} as is the case 
with edge states in phosphorene nanoribbons.

\begin{figure} \centering
	\includegraphics[width=8.6cm]{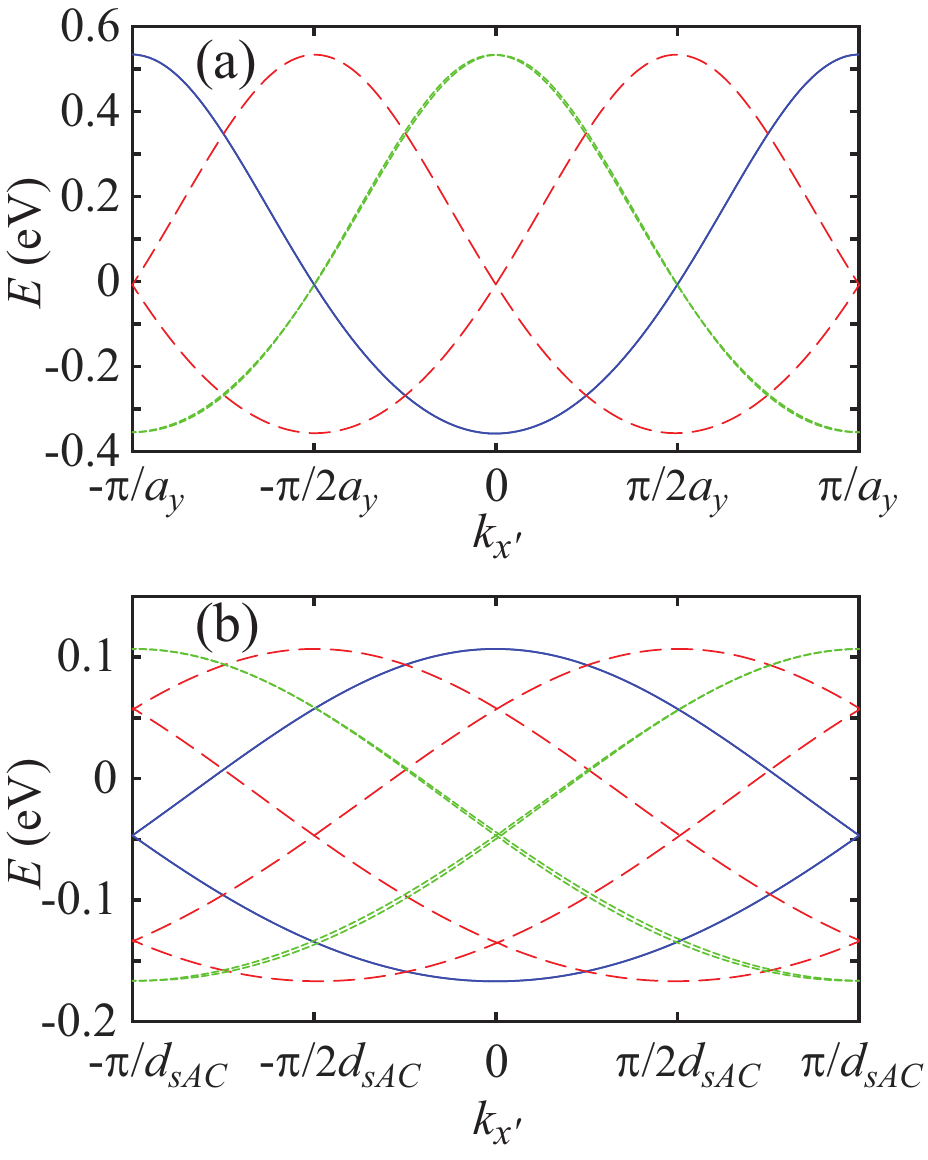}
	\caption{The quasi-flat bands for (a) $N_{ZZ}$=46 and (b) $N_{sAC}=74$ nanoribbon for $B=0$ (blue solid curves), 
	$B=B_{cr}/2$ (red dashed curves) and $B=B_{cr}$ (green dotted curves).}
	\label{fig8}
\end{figure}

In Fig.~\ref{fig8} we take a closer look at QFBs in (a) $N_{ZZ}=46$ and (b) $N_{sAC}=74$ under a range of magnetic field values.
In the absence of magnetic field, the bands are degenerate for both, ZZ and sAC, as depicted by solid blue curves in Fig.~\ref{fig8}.
It is clear that the QFBs get progressively shifted in opposite directions with increasing magnetic field (see red dashed lines in Fig.~\ref{fig8}). 
An explanation for this behavior is very similar to the one given in Ref.~[\onlinecite{grujic15}]; exponentially 
localized states "sample" the vector potential only in a small area where it is effectively constant, and 
therefore the effect of magnetic fields amounts only to a phase shift. Additionally, the two QFB states are 
localized at opposite edges (at $y'=\pm W/2$, for a ribbon of width $W$), where the local vector potential 
$\mathbf{A}=B\left(-y,0,0\right)$ has different signs, which explains the opposite shifts.

A quantitative description can be obtained by employing the principle of minimal coupling 
$k_{x'}\rightarrow k_{x'}+\frac{e}{\hbar}A_{x'}$. In other words, the bands get shifted left and 
right by $\frac{eBW}{2\hbar}$, depending on which edge they are localized at. We have confirmed that this is indeed a very good approximation for smaller fields. This also suggests that when $\pi/d=eB_{cr}W/2\hbar$, where $d$ is the unit cell length, QFBs get completely "out of phase", so that maxima and minima interchange places (see green dotted curves in Fig.~\ref{fig8}). This critical magnetic field is then given as 
\begin{equation}
B_{cr}=\frac{2\pi\hbar}{edW},
\end{equation}
which in fact means that the magnetic flux through the unit cell ($\Phi_{cr}=B_{cr}dW$) is equal to one flux quantum ($\Phi_0=h/e$). 

The exact expressions for the critical magnetic field in ZZ and sAC ribbons reads
\begin{eqnarray}
	B_{cr}^{ZZ}=\frac{\Phi_0}{(N_{ZZ}-1)P_{hex}},\\
	B_{cr}^{sAC}=\frac{\Phi_0}{2(N_{sAC}-2)P_{hex}},
\end{eqnarray}
respectively. Here, $P_{hex}=a_rc_r/2$ is the area of plaquet projection into the $x,y$ plane.
The green dotted curves in Fig.~\ref{fig8} show the band structure at $B_{cr}$. However, these two bands are 
not degenerate and thus not exactly in opposition with respect to the solid blue curves. There are two reasons for this, 
both having to do with the fact that edge states have some finite spread towards the ribbon center. On the one hand, 
this means that the edge states effectively experience somewhat smaller ribbon widths (thus increasing true $B_{cr}$). 
On the other hand, for larger magnetic fields the vector potential has a stronger spatial variation, so that the simple 
picture of phase shifts (assuming relatively constant $\mathbf{A}$ in the narrow space occupied by the edge state) 
gradually losses validity.

\section{Summary and Conclusions}

In summary, we derived a new Hamiltonian to describe the electronic structure of phosphorene nanoribbons 
with arbitrary edges in transverse electric and perpendicular magnetic fields. We found 
that when a magnetic field is turned on the states of positive and negative momenta
split to the opposite sides of the nanoribbon. An analytical expression for the minimal magnetic field 
when the bands become flat is obtained. The boundaries of these flat segments in momentum space 
are also described by analytical functions. We show that both the minimal field and the extension of 
the flat bands depend on the type of nanoribbon edge. Furthermore, when magnetic field increases, 
the Landau levels spectrum emerges first in an AC nanoribbon and last in the 
ZZ nanoribbon of equal width. Thus, the transversal confinement of electrons is
found to be weakest in the ZZ nanoribbons and strongest in the AC nanoribbons. Moreover, an application
of in-plane electric field causes the band gap to decrease and turns the previously flat
segments into ranges of linear variation of the energy spectra in momentum space. We found that the
electric field gives rise to shifts of crossings toward the center of the phosphorene Brillouin zone. 
For all ribbon types we found good agreement between the numerical results obtained by means of 
the tight binding and the results derived from an analytical model. Finally, the analytical expression 
for the critical magnetic when the edge states are localized at opposite edges acquire counter phases is 
derived for the cases of ZZ and sAC nanoribbons.

\begin{acknowledgments}This work was supported by the Serbian Ministry of Education, Science and Technological Development, and the Flemish Science Foundation (FWO-Vl).
\end{acknowledgments}

\end{document}